\documentclass[a4paper,10pt]{article}

\usepackage{a4wide}
\usepackage{graphicx}
\usepackage{xcolor}
\usepackage{amsmath}
\DeclareUnicodeCharacter{2009}{\,} 

\begin{document}

\title{Single-mode laser guiding in non-parabolic plasma channels for high-energy electron acceleration}

\author{Zsolt L\'ecz$^{1,2*}$, Szilárd Majorosi$^{1}$, Nasr A. M. Hafz$^{1,2}$}

\maketitle

\textit{1 ELI-ALPS, ELI-HU Non-Profit Ltd., Wolfgang Sandner utca 3., Szeged, H-6728, Hungary \\
2 Doctoral School of Physics, Faculty of Science and Informatics, University of Szeged, 9 Dóm tér, H-6720 Szeged, Hungary} 

$^*$Corresponding author: zsolt.lecz@eli-alps.hu

\hspace{10mm}

{\bf Abstract}
The discovery of laser wakefield acceleration in gaseous plasma was a major milestone that could lead to a significant reduction of size and cost of large electron accelerators. For higher-energy laser-driven electron acceleration guiding plasma channels were proposed, which are matched to the laser pulse parameters. For guiding a Gaussian beam, a parabolic density profile is needed, which is difficult to realize experimentally. The realistic channel profiles can be described by higher order polynomial functions which are not optimal for guiding due to the development of undesired distortions in the laser intensity envelope. However, here we show that for non-parabolic plasma channels well-defined matching conditions exist, which we call mode matching. This leads to the guiding of the fundamental mode only in the acceleration regime, where the plasma electron density is modulated by the high-intensity laser pulse. In this way we eliminate two deteriorating factors of laser wakefield acceleration, namely the mode dispersion and energy leakage. We apply this new matching condition for single-mode guiding in quasi-3D simulations to show that 10 GeV energies can be reached in a distance of less than 15 cm.


\section{Introduction}

High-energy RF-based (radio-frequency) particle accelerators delivering beams of charged particles have been of crucial importance in industry, science and technology for their applications, for example, in cancer therapy, material science, photon science and in discoveries of fundamental particles in nature \cite{book1, applic1}. A good example for an operational large-size machine is the 3.4 km-long European X-ray Free Electron Laser (XFEL), which is driven by 17.5 GeV electron beams from a 1.7 km-long superconducting LINAC (linear accelerator). However, the cost of multi-km long accelerators is very high, especially, when the required particle energies are well-beyond one gigaelectronvolt (GeV) \cite{highCost}. So, there is an important motivation for physicists and engineers to create alternative techniques to
down-size particle accelerators in order to make them more affordable. Since the invention of the chirped-pulse-amplification technique \cite{CPA}, the laser pulses from table-top laser systems can be sufficiently intense to drive plasma wakefields \cite{LWFA} in gas (such as Hydrogen or Helium) targets that can be used to accelerate electrons to relativistic energies \cite{100MeV1, 100MeV2} over extremely short distances. This is called laser wakefield acceleration (LWFA), a laser-based acceleration scheme involving CPA laser pulse as driver and underdense gaseous plasma medium. Laser wakefield accelerated GeV-class electron bunches by 100s of TW laser pulses have been demonstrated for more than a decade carrying 10s to 100s of pC total charge \cite{gevcm, Banerjee,gevNasr, nearGeV,2gev, selfTrunc}. The quality of these electron beams were sufficiently good for some applications, such as betatron X-ray radiation imaging and micro-CT (computerized tomography) of various objects \cite{tomo}. On the other hand, there are continuous efforts to bring up the quality of the GeV beams from LWFAs to meet more application requirements. It is also worth mentioning that there has been great efforts in the development of electron-beam driven plasma wakefield accelerators \cite{Edriven}, but in that case the energy conversion efficiency is still too low.

LWFAs have shown a tremendous advancement \cite{reviewEsarey, reviewHidding}, thanks to the availability of commercial and in-house-built table-top terawatt and petawatt laser systems around the world \cite{ELIALPSPW, 10PW}. The ever-increasing laser peak power led scientists to approach the 10 GeV energy level within 20 cm acceleration distance \cite{leemans, gonsales} in the plasma. Electron bunches with such high energy, which are naturally synchronized to the driving laser pulses, are important for revealing fundamental QED processes, such as the photon-recoil effect \cite{PRXQeD1, PRXQeD2}, for linear colliders \cite{lincoll}, and for driving XFEL-s.

\begin{figure}[h]
\centering
\includegraphics[trim= 0 0mm 0 0mm,width=0.91\textwidth]{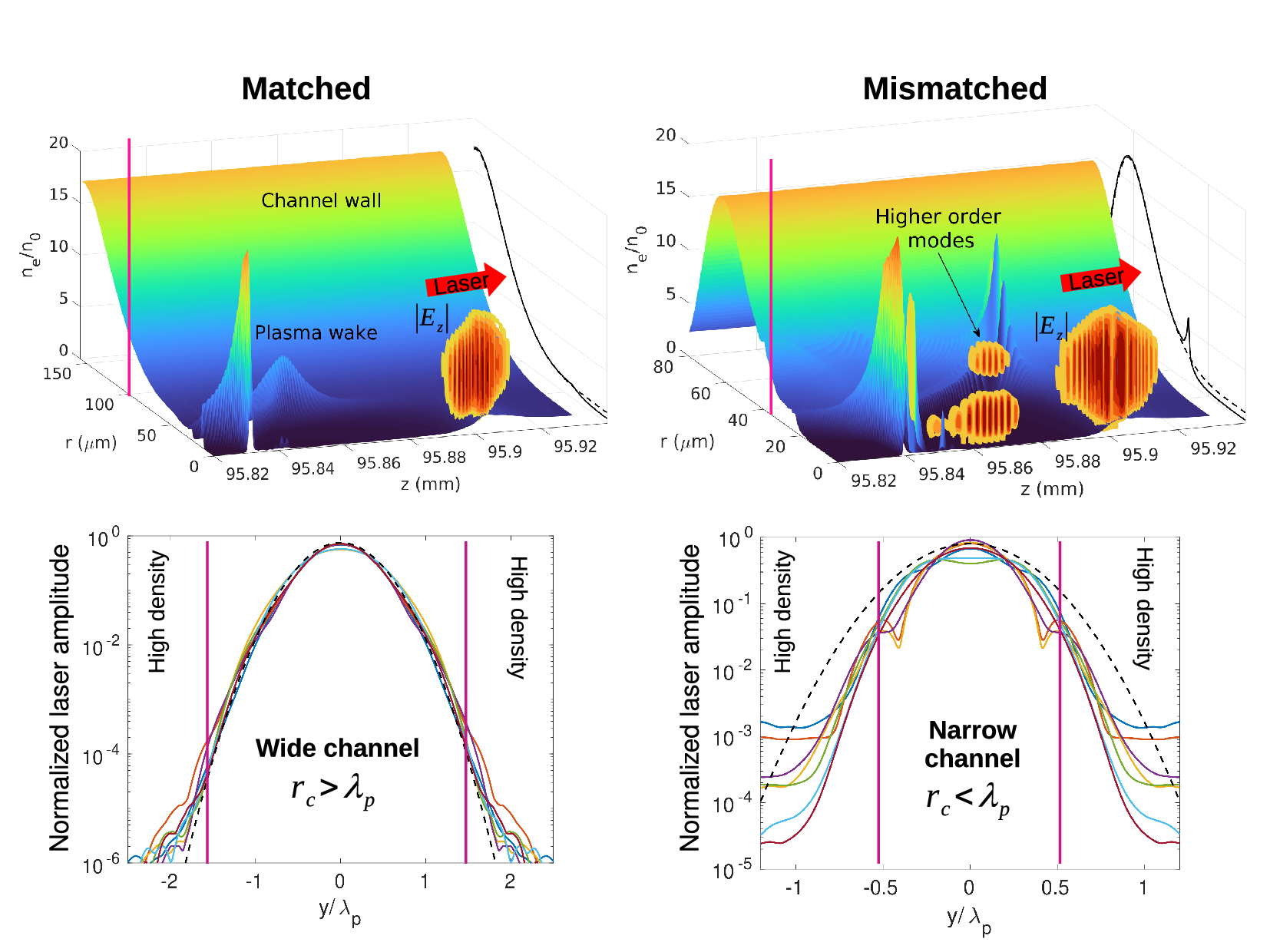}

\caption{{\bf Concept of the single-mode guiding for a relativistic-intensity Gaussian laser pulse:} We show the matched case on the {\it left} and the mismatched case is on the {\it right}. The figures in the top show the plasma density after $\sim$10 cm of propagation overlaid with the vertical cross-section of the linearly polarized laser pulse as orange-red colors. The longitudinal component of the laser field is used to represent the higher order modes and the main pulse with the same color code. On the bottom we show the z-averaged (within the full width at half maximum) absolute transverse field component at different time instances (solid curves) with the initial Gaussian-profile (dashed black curves) and we indicate the position at the channel radius ($|y|=r_c$) with vertical lines. Here $\lambda_p=66 \mu$m. In the mismatched case the laser field contains higher order modes with significant amplitude that propagate behind the main pulse due to mode dispersion, while the laser pulse preserves its Gaussian-profile in the matched case. }
\label{firstFig1}
\end{figure}

Several methods have been developed to extend the acceleration length and to increase the electron energy. External guiding has been applied in the form of discharge capillary waveguides \cite{gevcm, leemans, gonsales}. The most recent and most reliable channels are formed in elongated gas jets by means of hydrodynamic expansion followed by the optical field ionization of the gas medium, so-called HOFI channels \cite{tunableHofi, OFI, Axiparabola}, that can provide a meter-scale acceleration length \cite{meterScale} and operation at kilohertz repetition rate \cite{kilohertz}. These plasma channels are highly flexible and can be easily matched to the laser focusing optics by tuning the delay time between the generating beam and main driver pulse.

Even in the case of improved laser guiding the acceleration is limited by the pump energy depletion and electron dephasing \cite{depahsing}. In the case of low-density plasma channels, the latter is the critical negative effect that strongly limits the electron energy gain. By applying spatio-temporal shaping of the laser pulses, one can circumvent this limitation \cite{debus, dephaseless, multipulseDeph}, but these advanced methods have not been tested with plasma channels yet. In the case of a guided laser pulse, another effect, which is the mode dispersion \cite{clark2000}, can also ruin the acceleration efficiency. This is caused by the fact that the realistic channel density profile strongly deviates from the ideal parabolic shape, and higher-order Laguerre-Gaussian modes appear. In recent experiments, where the laser spot and channel entrance had similar diameters, these higher order modes were present in the channel \cite{OFI, selfInjHOFI, modeEvol, 10GeV}, that caused strong envelope oscillation and energy leakage in the first several centimeters of the propagation. 

In this work, we consider non-parabolic radial density profiles, very similar to those existing in hydrodynamic plasma channels \cite{tunableHofi, OFI}. We propose the concept of a wide channel, where the channel diameter is much larger than the laser waist and it is larger than the plasma wavelength, allowing the propagation of the fundamental Gaussian laser mode only, thus the mode dispersion or mode beating effects can be almost completely eliminated. This regime of laser guiding relies on the self-generated channel density profile that is modulated by the laser's ponderomotive force. The idea is that, instead of using a parabolic channel density profile we consider a more realistic non-parabolic shape that is modulated by the laser pulse such that the plasma susceptibility, seen by laser pulse, acquires a parabolic shape. The self-modulated channel density and the increased relativistic mass of electrons establishes a condition which is ideal for stable single-mode propagation of an intense laser pulse in arbitrary channels, which is usually highly nonlinear and difficult to control in uniform plasma. Our scheme is presented in Fig. \ref{firstFig1}, where two examples are shown: the left part corresponds to the ideal guiding (wide channel) and the right part present a non-ideal, mismatched case, where the laser envelope deviates significantly from the fundamental mode. In the case of matched guiding, the plasma wavelength is smaller than the channel radius, which is indicated by the vertical dashed lines in the lower pictures. We also show that the electron injection into the accelerating wake occurs more naturally in wide channels, whereas in other cases it is not always the case \cite{selfInjHOFI}.

\section{Single-mode guiding of intense Gaussian laser beams}

The fundamental mode of a laser beam is guided without the development of higher-order modes (HOMs) if the plasma susceptibility has a parabolic radial profile. It can be proven easily considering the reduced wave equation of a monochromatic electric field $E_x={\tilde E} e^{-i\omega( t - z/c ) }$, with angular frequency $\omega$, and considering cylindrical symmetry:

\begin{equation}\label{eq:helm}
\frac{1}{r}\frac{\partial }{\partial r} \left( r \frac{\partial {\tilde E}}{\partial r} \right)
+ 2i\frac{\omega}{c}\frac{\partial {\tilde E}}{\partial z}=
\frac{\omega^2}{c^2} \chi(r)  {\tilde E},
\end{equation} 
where ${\tilde E}=E_0\exp(-r^2/w^2 - i\theta)$ is the complex envelope of a Gaussian beam that contains the information about the transverse phase profile (wavefront). We also introduce the relativistic plasma susceptibility expressed as
\begin{equation}\label{eq:chih}
\frac{\omega^2}{c^2} \chi(r) = 
\frac{e^2}{\varepsilon_0 c^2 m_e } \frac{n_e (r)}{\gamma}  = 
\frac{k_p^2}{ \gamma} F(r),
\end{equation} 
where $n_e(r) = n_0 F(r)$ is the radial profile of the plasma density seen by the laser pulse, and $n_0$ is the electron density on the axis of the channel ($r = 0$), which defines the value of the plasma wave number $k_p=2\pi/\lambda_p$ and $\gamma=\sqrt{1+a_0^2|{\tilde E(r)}/E_0|^2/2}$ with $a_0=eE_0/(m_e c\omega)$. The plasma wavelength is defined as $\lambda_p=2\pi c/\omega_p$, $\omega_p=(e^2n_0/m_e\epsilon_0)^{1/2}$. We also note here that the relation between $\chi$ and the index of refraction ($\eta$) is simply: $\chi=1-\eta^2=n_0/(n_c\gamma)$, where $n_c=\omega^2m_e\varepsilon_0/e^2$ is the critical density.

Inserting the formula of ${\tilde E}$ in Eq. (\ref{eq:helm}) leads to the evolution equation for the phase shift at the beginning of the interaction:

\begin{equation}\label{eq:theta}
\frac{2\omega}{c^2} \frac{\partial \theta}{\partial \tau} = \frac{k_p^2}{\gamma} F(r)  - \frac{4}{w^2} - \left(\frac{\partial \theta}{\partial r} \right)^2 -\frac{4 r^2}{w^4}.
\end{equation} 
If $\gamma=1$ and  $\partial \theta/\partial r=0$ initially, one can easily see that the phase shift will be constant in $r$ if $F=F_{m}=1+4r^2/(k_p^2w^4)$, which is the condition for perfect guiding \cite{EsareyGuide}. This leads to the conventional matching condition, where the parabolic channel radius is expressed as: $r_c^2= k_p^2w_0^4/4$. In general, in vacuum ($F=0$), the phase of a Gaussian laser beam is a parabolic function of radius, $\theta = \theta(\tau, r^2)$. It follows from Eq. (\ref{eq:theta}), that in plasma $\theta$ remains a purely parabolic function along the propagation if $F(r)/\gamma=1+Cr^2$, where $C$ is constant in $r$, because $\partial \theta/\partial r$ depends linearly on $r$. This ensures that the guided laser pulse preserves its initial Gaussian radial shape.

Therefore, the basic requirement to avoid the development of higher order modes is the existence of a parabolic electric susceptibility in the radial direction, i.e. perpendicular to the laser propagation direction. In the case of low intensity (non-relativistic) laser pulses this requirement is fulfilled by a parabolic density profile. The actual density profile seen by a high-intensity laser pulse is modulated by the laser field: 

\begin{equation}\label{eq:densfull}
n_e(r) = n_0 F(r)= n_0\left(f(r)+\delta n \right),
\end{equation}
where $f(r)=1+\epsilon (r/r_c)^b$ is the unperturbed plasma channel profile and $\delta n$ is the ponderomotive density modulation which is different at each longitudinal position within the laser pulse. This leads to asynchronous oscillation of the laser spot, because of the non-uniform index of refraction along the laser pulse. This can be compensated by defining a laser pulse with varying laser spot size along its axis, which is called super matching \cite{benedetti2}. In this work, we do not look for conditions of constant spot size, rather we show that the self-consistent plasma susceptibility preserves the quasi-parabolic shape when envelope oscillation is allowed.

One can recognize that in Eq. (\ref{eq:helm}) the radial coordinates can be normalized with $k_p$ (using the identity Eq. (\ref{eq:chih})), thus from now on, $\chi=n_e/(n_c\gamma)$ is replaced by ${\omega_p^2}/{\omega^2} \chi=n_e/(n_0\gamma)$. We can denote $\chi=F/\gamma=n_e(r)/(n_0\gamma(r))$ as the modified and $\chi_{opt}=\chi_0+4 \kappa r^2/(k_pw_0^2)^2$ as the ideal susceptibility profile in the plasma channel for optimal (single-mode) guiding. Here $\chi_0=n_e(0)/(n_0\gamma_0)$, with $\gamma_0=\gamma(r=0)$, and $\kappa$ is a free parameter, that expresses the focusing strength of the modified plasma channel. With the notations of standard guiding theory \cite{EsareyGuide} this quantity is nothing else but the normalized channel depth $\kappa=\Delta n/\Delta n_{cr}$, where $\Delta n=n_e(w_0)-n_0$ and $\Delta n_{cr}=(\pi r_e w_0^2)^{-1}$, $r_e$ is the classical electron radius. In principle, if $\kappa=1$, then the laser pulse is guided with constant spot size, but it is never the case, because it varies along the laser pulse. When $\kappa>1$ the laser spot size oscillates between $w_0$ and $w_0/\sqrt{\kappa}$. In the next section we seek for laser parameters that ensures $\chi\approx \chi_0+4 \kappa r^2/(k_pw_0^2)^2$ for finite pulse durations.

\section{Existence of mode matching at relativistic intensities}

Laser propagation in plasma channels is a highly nonlinear multi-parametric problem, which depends on the following physical quantities: axial electron density ($n_0$), laser field amplitude ($a_0$), laser waist radius ($w_0$), pulse duration ($t_L$), plasma channel radius ($r_c$) and it depends on the channel's density profile as well. In this work we tighten this vast space of parameters and restrict our analysis to $ct_L\approx 0.3\lambda_p$, $r_c>w_0$ and $w_0<\lambda_p$. It is shown later that in this parameter-range the self-generated $\chi$ can be well fitted to a parabolic function. Furthermore, from the standard theory of nonlinear wakefields \cite{wbMori} it is known that longitudinal density modulation enhances the spectral modulation of the laser pulse, which in turn enhances dephasing. In order to suppress this effect the laser field amplitude has to be small, or the laser waist must be smaller than the plasma wavelength \cite{densModPlasma}: $w_0<\lambda_p$.

We consider a non-parabolic plasma channel, that is more realistic and comparable with density profiles obtained in HOFI channels used in recent experiments. See Eq. (\ref{eq:densprof}) in the Appendix for a formula of a guiding plasma channel which is created by optical field ionization. Exact analytical description of the wakefield structure is not possible in the high-intensity short pulse regime, but we show that $\delta n$, according to the simulations, can be approximated by the following function:

\begin{equation}\label{eq:densmod}
\delta n  \approx - g(\xi)\frac{4a_0^2 \exp(-2s^2)}{k_p^2w_0^2\sqrt{1+(a_0^2/2)\exp(-2s^2)}},
\end{equation}
where $s=r/w_0$. We note here that this function is very similar to the one obtained by evaluating to the widely used expression for long pulses \cite{bychenkov2018, EsareyGuide}: $\delta n = k_p^{-2}\nabla_r^2 \gamma$. In this work we use a numerical factor $g$ in order to take into account the longitudinal variation of the plasma density: $g\approx -0.5$ in the front of the laser pulse and $g\approx 1$ in the back, while in the center it is close to zero. This numerical factor, in the first half of the laser pulse, is close to the value that one can get after evaluating the integral in Eq. (119) in Ref. \cite{EsareyGuide} or Eq. (9) in Ref \cite{benedetti} considering a pulse length around one third of the plasma wavelength. From Eq. (\ref{eq:densmod}) it follows that in order to avoid full blow-out regime ($n_e(0)>0$) the spot size should satisfy $k_pw_0>\sqrt{2}a_0/(1+a_0^2/2)^{1/4}$ for $g\approx 0.5$, meaning that $w_0>0.4\lambda_p$ for $a_0<3$, which holds in our parameter-space.

\begin{figure}[h]
\centering
\includegraphics[trim= 0 5mm 0 5mm,width=0.99\textwidth]{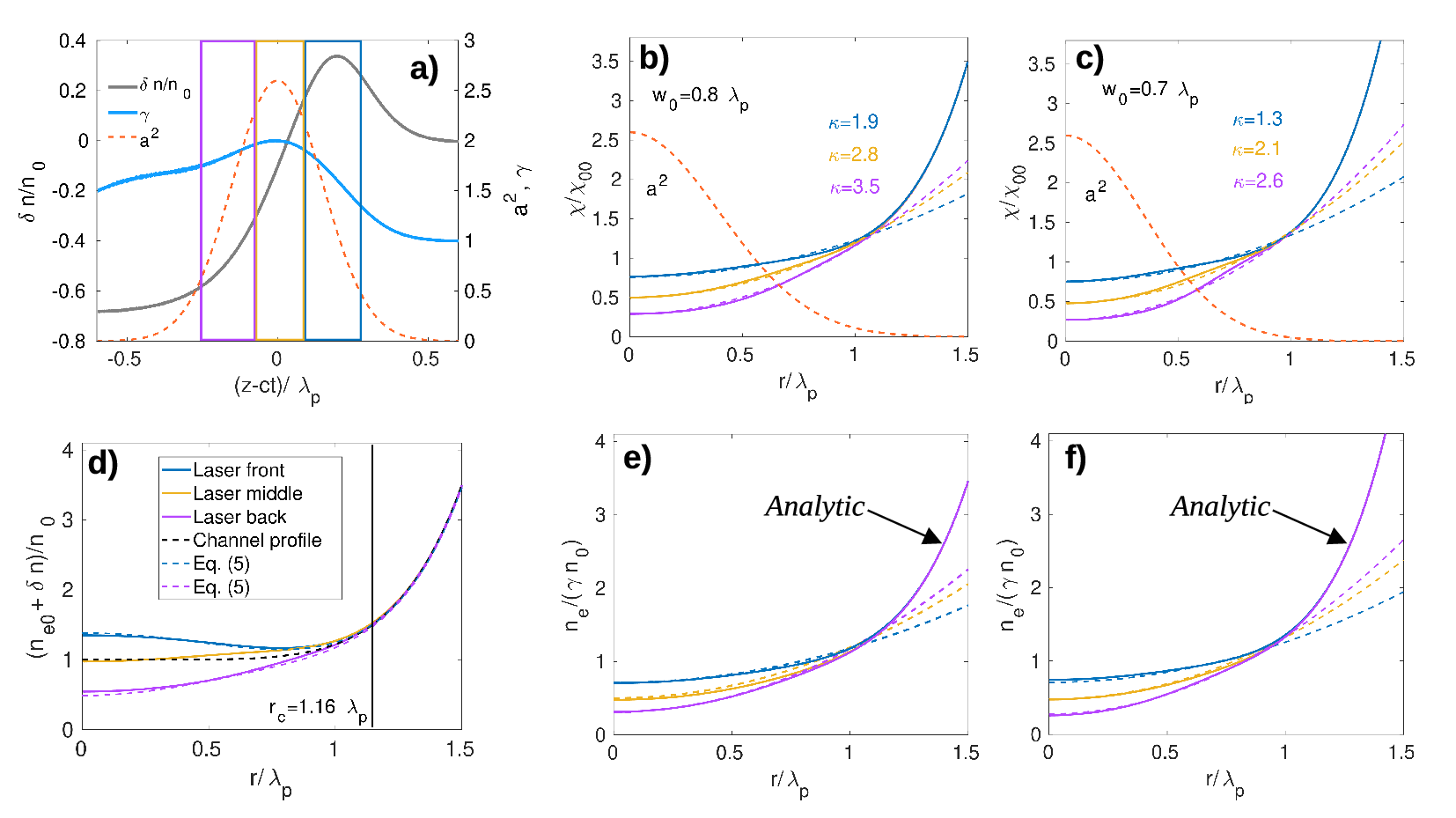}

\caption{{\bf Plasma density and susceptibility inside the laser pulse:} (a) Longitudinal profile of the plasma wave is presented for $w_0=0.8\lambda_p$. The radial profile of $\chi$ is presented in (b) in the front, middle and black of the laser pulse, averaged longitudinally in the zones indicated by the rectangles in (a). The same is shown in (c), but for $w_0=0.7\lambda_p$. The radial density profiles (corresponding to (a)) are presented in (d), which are compared to the analytical expression of $\delta n$ given in Eq. (\ref{eq:densmod}). The dashed curves in (b,c,e,f) correspond to the parabolic function ($\chi_{opt}$) with different kappa values, shown by the same colors. In all cases the ratio of channel radius to laser waist is the same: $r_c/w_0=1.45$ and $b=6$. }
\label{explainFig}
\end{figure}

For easier presentation and better clarity we use the laser envelope solver, available in SMILEI \cite{SMILEI}, to illustrate the density modulation generated by the ponderomotive force in a wide plasma channel. In this code the susceptibility is saved as a field quantity, therefore the comparison with our analytical modeling is straightforward. The grid size is 100 nm in the $z$ direction and 200 nm radially. In each cell 4 macro-particles are placed and only the zeroth (fundamental) azimuthal mode is used. 

In Fig. \ref{explainFig} the structure of the wakefield is presented for two laser waist radii, but for constant relative channel radius: $r_c/w_0=1.45$. One can see that in the front of the laser pulse the density modulation is positive, while in the back it is negative, therefore a zone must exist where $\delta n\approx 0$. For the pulse duration used in our work the zero density modulation happens to be close to the center of the laser pulse, which is very important for the analytical modeling, presented later. More importantly the plasma susceptibility agrees almost perfectly with the parabolic shape, desired for single-mode guiding, in all zones of the laser pulse. This is a crucial property of the wakefield, which exist when $a_0$ is large enough, because despite the positive density pile up the susceptibility is still convex and parabolic due to the large $\gamma$ of electrons. We note here that for larger values of $r_c/w_0$ (when $w_0<0.5\lambda_p$) the susceptibility starts to deviate from the ideal parabolic shape, therefore very strong self-focusing (large $\kappa$) must be avoided.

From Fig. \ref{explainFig} it is clear that different longitudinal segments of the laser pulse are guided differently, but the high order modes (HOM) can be suppressed everywhere if the parabolic shape of $\chi$ is preserved during the propagation. This is what we see in simulations, but in order to gain more insights we apply the method of source-dependent expansion (SDE, see Appendix) \cite{SDE} to analyze the laser beam evolution in the plasma channel, using an analytical description of the self-consistent susceptibility, shown in Fig. \ref{explainFig}(e,f). First, we consider $\delta n=0$, which is valid in the central part of the laser pulse, thus the susceptibility will be defined only by the laser envelope via $\chi= f(r)/\gamma(r)$.

\begin{figure}[h]
\centering
\includegraphics[trim= 9mm 0mm 9mm 0mm,width=0.95\textwidth]{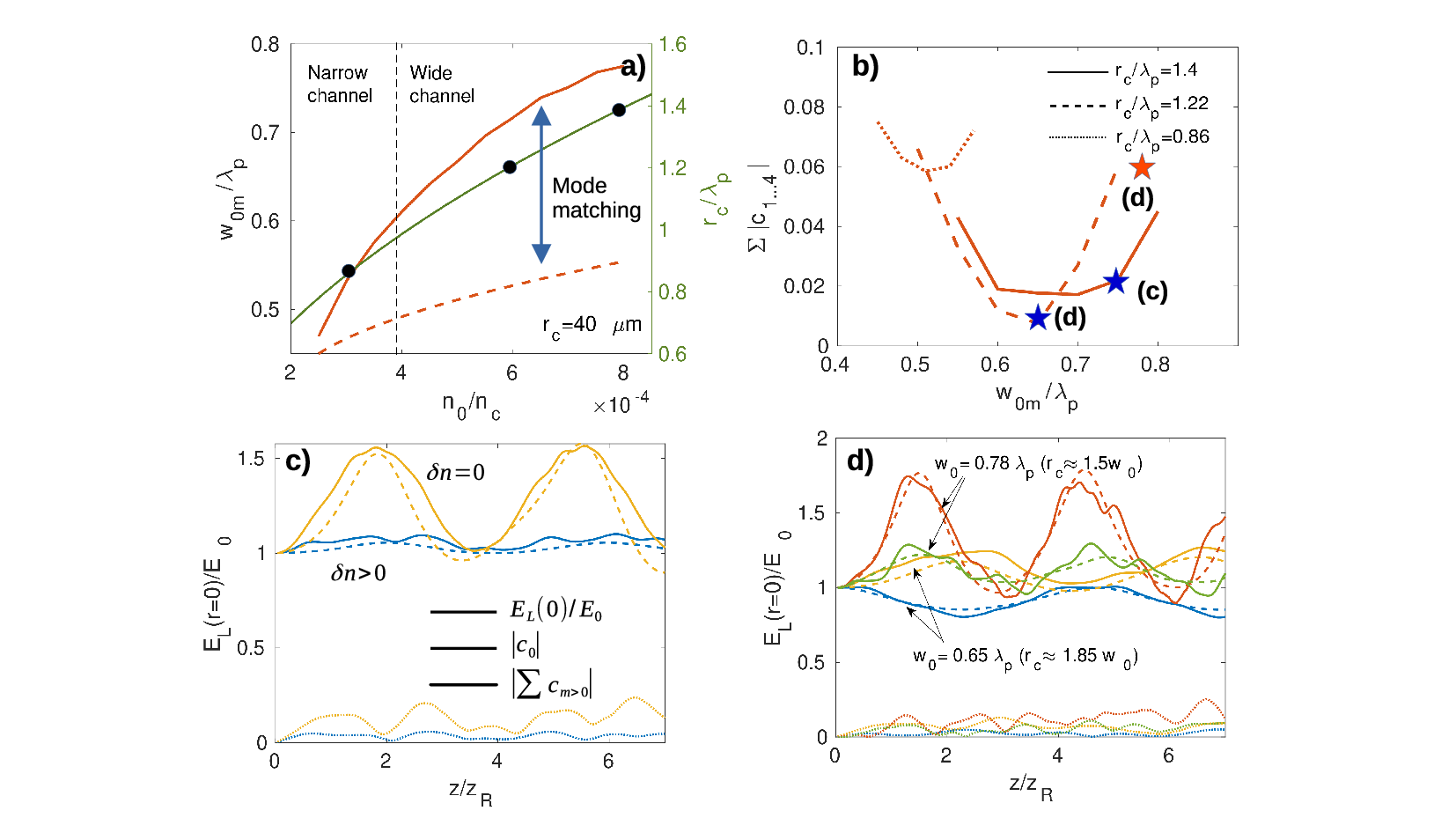}

\caption{{\bf Parameter-scan for high-order mode suppression:} (a) The upper limit of optimal waist radius (full lines) is shown for mode matching with $a_0=2.6$ as a function of plasma density. There is an interval between the full and dashed lines, where mode-matching is possible. In (b) the sum of the higher order mode coefficients ($c_m$) are shown as a function of laser spot size for different normalized channel radii (corresponding to different plasma densities, shown by the black dots in (a)). The absolute values of individual modes (normalized to $a_0$) are shown in (c) and (d) for different density and $w_{0}$ values. The amplitude of high-order modes is very small in all cases.  }
\label{SDEscan}
\end{figure}

Using the SDE method we can monitor the evolution of HOMs within the laser beam as it propagates in the channel and we aim to minimize the amplitude of those modes. For this purpose an extensive parameter-scan has been performed where the laser field amplitude ($a_0=2.6$) and channel radius ($r_c=40\,\mu$m) were fixed and the initial laser waist radius is varied. For a given axial density value ($2\times 10^{-4}<n_0/n_c<10^{-3}$) we look for the values of $w_{0m}$, where the amplitude of all high order modes, averaged in time, is less than 10 \% of the fundamental mode over many Rayleigh lengths. The resulting matched laser spot radii are shown in Fig. \ref{SDEscan}(a) for a steep channel wall ($b=6$). One can see that there is a finite $w_0$ interval where this condition is met and the width of the interval depends on the plasma density. These intervals are shown in Fig. \ref{SDEscan}(b) for three density values, which correspond to three different normalized channel radii. 

The evolution of the fundamental mode and the contribution of all other modes are presented in Fig. \ref{SDEscan}(c,d) for parameters indicated by the stars in Fig. \ref{SDEscan}(c). The resultant amplitude of HOMs is less than 20 \% in all cases, which means that more than 95 \% of the laser energy maintained in the fundamental mode. It is important to see what happens in the front of the laser pulse, which means a positive $\delta n$ in the source term of the envelope equation. This is presented by the blue curves in Fig. \ref{SDEscan}(c,d), which  shows a distinct behavior, but the amplitude of HOMs is even smaller. In the previous section it was shown that for a Gaussian mode $E_{L,max}/E_0=\sqrt{\kappa}$, which means that in Fig. \ref{SDEscan}(c) $\kappa<1$ in the front and $\kappa\approx 1.7$ in the middle of the pulse, while in Fig. \ref{SDEscan}(d) these values are $\kappa\approx 1$ and $\kappa\approx 2.4$, respectively, for the smaller laser spot. The case of the larger spot roughly corresponds to the case shown in Fig. \ref{explainFig}(b), where $\kappa=2.8$ was identified in the center of the pulse, corresponding to $E_{L,max}/E_0\approx 1.7$, which agrees with the modeling in Fig. \ref{SDEscan}(d). The SDE method cannot provide valid modeling in the back of the laser pulse, because $\delta n$ does not depend solely on the local laser field, but its shape is mostly governed by the front of the laser pulse when $w_0$ in the back becomes  smaller than $w_0$ in the front.

From this analysis we see that mode matching is possible in wide channels, but for coherent oscillation of the laser envelope $\kappa>2$ is required at the center of the pulse. For smaller values the front side will diffract ($\kappa<1$), as it is seen in Fig. \ref{SDEscan}(c), which causes very different oscillation periods of the laser spot at the front and back sides of the pulse. One solution to realize guiding with constant spot size is called super-matching, which is presented in Ref. \cite{benedetti2}. On the other hand, $\kappa$ cannot be too large, because that would result in strong self-focusing leading zero plasma density in the back size of the laser pulse (blow-out regime).

\begin{figure}[h]
\centering
\includegraphics[trim= 0 0mm 0 0mm ,width=0.8\textwidth]{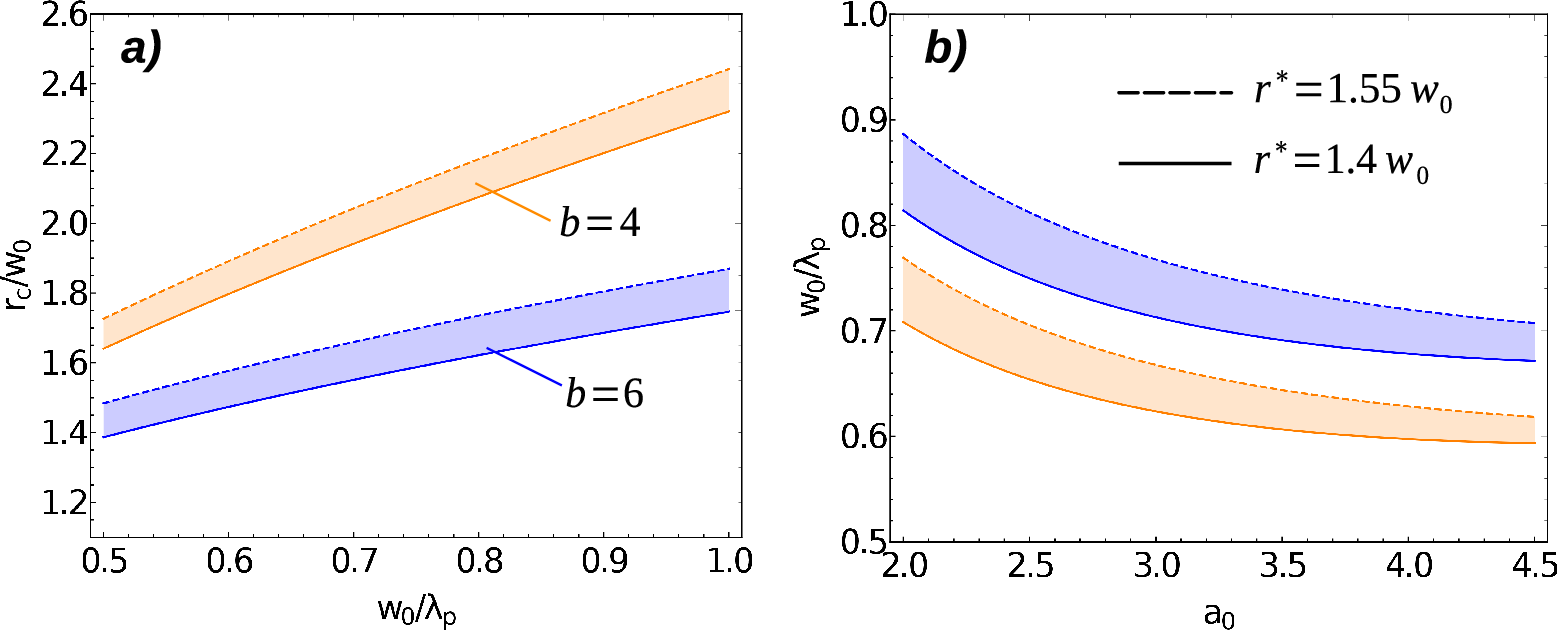}

\caption{{\bf Parameters for single-mode guiding:} (a) Matched channel radius is shown as a function of laser spot size according to Eq (\ref{eq:rc}). (b) Matched laser waist radius amplitude versus normalized laser field amplitude calculated with Eq. (\ref{eq:w0m}). Here $\kappa=2$. }
\label{resultFig}
\end{figure}

Finding the mode-matching conditions analytically would be extremely challenging if a precise $\delta n$ has to be considered, but from the wakefield structure and guiding analysis presented above it is evident that $\delta n=0$ can be chosen and $2<\kappa<3$ has to be considered for stable guiding. With this simplification the matched laser parameters can be found from the equality $\chi=\chi_{opt}$, leading to:

\begin{equation}\label{eq:fitting}
\frac{1+\epsilon(r/r_c)^b}{\gamma(r)}=\gamma_0^{-1}+\kappa \frac{4 r^2}{k_p^2 w_0^4},
\end{equation}
where $\epsilon$ and $b$ are known and can be taken from Table \ref{Table1}. The problem of finding the matching condition is reduced to a fitting problem where the parabolic shape has to be joined with the initial density profile at a position $r^*$. A logical choice for this fitting point is $r^*=1.5w_0$, because it is far enough from the axis to provide matching for most of the laser pulse's energy and it is also far enough from the region $r>r_c$, where the density gradient is higher than that of a parabolic function. Having a smooth transition between the parabolic zone and outer zone of the susceptibility requires $d\chi/dr=d\chi_{opt}/dr$ at $r=r^*$, which results in the matched channel radius for a given laser spot radius:

\begin{equation}\label{eq:rc}
r_c=r^*\left( \frac{8\kappa {s^*}^2}{b \epsilon k_p^2 w_0^2} \right)^{-1/b}
\end{equation}
where $s^*=r^*/w_0$ and $\gamma(r^*)\approx 1$ is assumed, i.e. the effect of $\gamma$ in the shape of the plasma channel is negligible at $r\ge r^*$. Inserting this expression in Eq. (\ref{eq:fitting}) leads to the matched condition for the laser pulse:

\begin{equation}\label{eq:w0m}
k_p^2w_0^2=4(\gamma_0 {s^*}^2\kappa/b) [b \gamma^* -2]/(\gamma_0-\gamma^*),
\end{equation}
where $\gamma^*=\gamma(r^*)$. Now Eq. (\ref{eq:w0m}) represents a new matching condition that ensures a clean propagation of the fundamental mode only, and it is different from the matching condition used to suppress the envelope oscillation in a parabolic channel, well-known in the literature \cite{EsareyGuide}. The resulting matched parameters are shown in Fig. \ref{resultFig}, where the effect of $r^*$ is also shown, which might look as an uncertainty in the modeling. However, varying $r^*$ or $w_0$ is equivalent to varying $\kappa$, therefore the region between the dashed and full lines in Fig. \ref{resultFig} can be considered as a degree of freedom in choosing the initial parameters. With other words, by choosing 10 \% smaller or larger laser spot in a given physical setup will always result in radial mode suppression, but the period and amplitude of the envelope oscillation will be different.

\begin{figure}[h]
\centering
\includegraphics[trim= 0 0mm 0 0mm,width=0.8\textwidth]{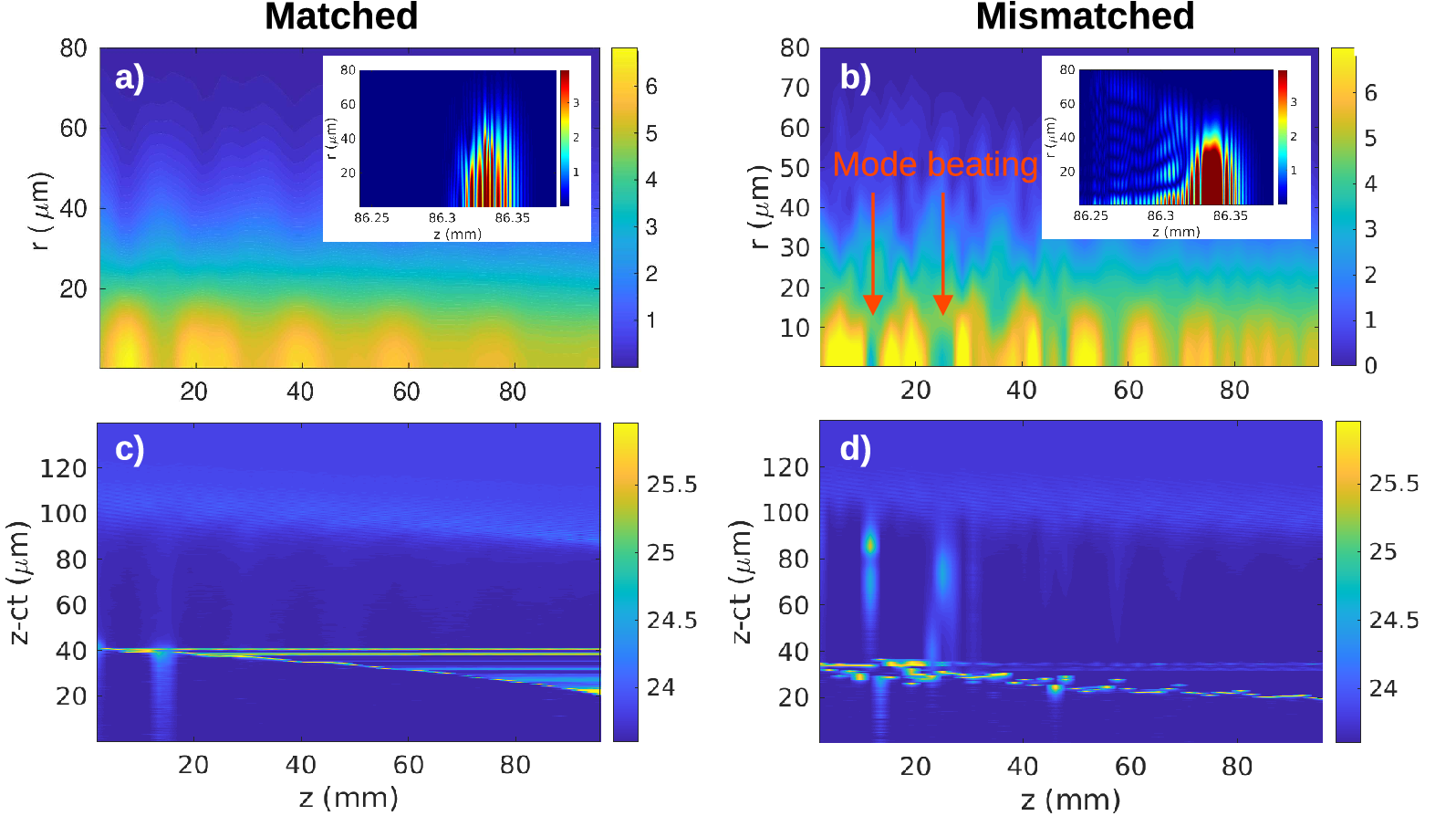}

\caption{{\bf Comparison between matched and mismatched guiding:} (a) and (b) shows the evolution of the radial profile of the axially averaged laser field amplitude (in units of TV/m). The insets show the cross-section of the laser pulse near the end of the channel. Strong mode dispersion in the mismatched case is clearly visible. In (c) and (d) the axial lineouts of the wake density in the first bubble are shown on logarithmic scale.  }
\label{Compare}
\end{figure}

For the numerical modeling of the long distance propagation we use FBPIC \cite{fbpic}, a quasi-3D simulation code, with the moving window feature. The velocity of the window is exactly the speed of light in vacuum. The simulation domain is represented by $3000\times800$ grid points, with grid spacing of 40 nm and 200 nm in the longitudinal and transverse dimensions, respectively. Two azimuthal modes are used and the number of macro-particles per cell is 64 ($N_z\times N_r\times N_{\theta}=2\times 4\times 8$). We assume pre-ionized gas and the ions are immobile, forming a neutralizing positive background. The radial plasma density profile is given by Eq. (\ref{eq:densprof}). The laser pulse has Gaussian temporal and spatial envelope, and it is focused at the entrance of the plasma channel.

The example shown in Fig. \ref{firstFig1} corresponds to the matched case for $b=4, w_0=0.6\lambda_p, a_0=2.5, \kappa=1.7$ and $r_c=1.1\lambda_p=1.8w_0$. We further illustrate the importance of guiding in a wide channel in Fig. \ref{Compare}, where two simulations are shown with exactly the same laser pulse, but the channel parameters are different. The laser parameters are $w_0=0.7\lambda_p=40\,\mu$m, $a_0=2.5$ ($\kappa=2$) and it has 22 J energy. In the matched case the channel radius is calculated using Eq. (\ref{eq:rc}) and it is $r_c=1.4\lambda_p=82\,\mu$m with $n_0=3.45\times10^{17}$ cm$^{-3}$, while in the mismatched case $r_c=w_0=0.54\lambda_p=40\,\mu$m with $n_0=2\times10^{17}$ cm$^{-3}$. In the matched case very smooth and regular envelope oscillation is observed, while in the mismatched case strong mode beating takes place, when the laser intensity maximum is not on the axis. This has a significant influence on the wakefield formation, shown in Fig \ref{Compare}(d). The spectra of generated electron bunches are shown in Fig. \ref{spectraComp} (for 22 J pulse energy).

\section{Periodic injection of plasma electrons into an oscillating bubble}

\begin{figure}[h]
\centering
\includegraphics[trim= 0 5mm 20mm 5mm,width=0.48\textwidth]{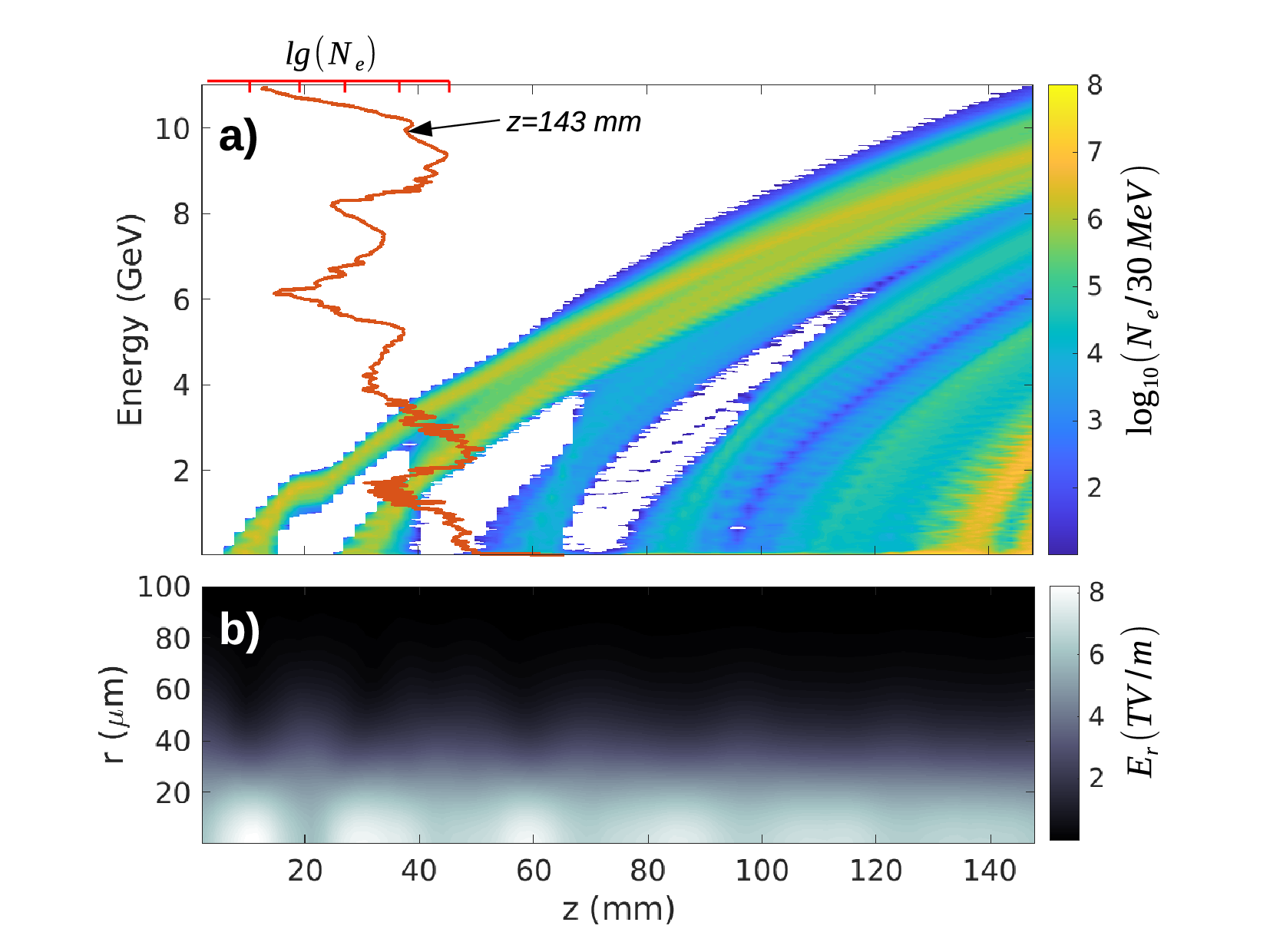}
\includegraphics[trim= 20mm 5mm 0 5mm,width=0.47\textwidth]{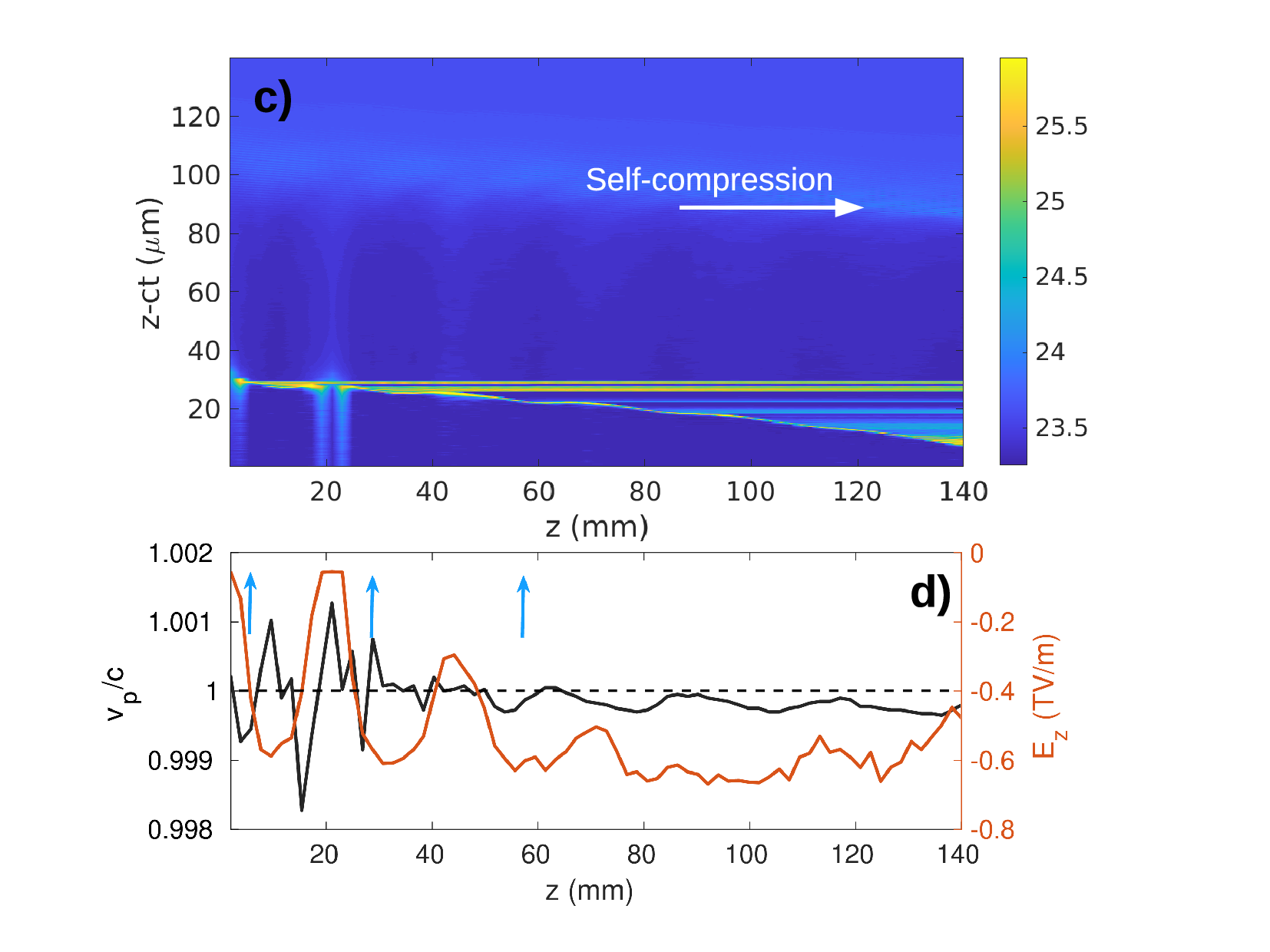}

\caption{{\bf Temporal evolution of the acceleration:} (a) Electron energy spectrum along the propagation distance. In (b) the axially averaged laser field modulus is shown, where the periodic focusing and defocusing of the laser pulse along the plasma channel is clearly visible. Each focusing triggers an injection event. The history of the axial density profile ($log_{10}[n_e]$ with units of $m^{-3}$) of the nonlinear wakefield is shown in (c), where the horizontal stripes correspond to the injected electron bunches moving with nearly the speed of light. (d) The peak electric field in the back of the bubble is shown by the red line, while the velocity of the node of the wave (phase velocity of the plasma wave) is shown by the black line. As indicated by the blue arrows, self-injection occurs, when the electric field is strong enough and the phase velocity is sufficiently low.  }
\label{secondFig}
\end{figure}

An intrinsic property of the proposed wide plasma channels is that the self-generated parabolic profile leads to self-focusing ($\kappa>1$), because the radial steepness of the refractive index is larger than the one required for perfect guiding. This can result in self-injection of the electrons when certain conditions are met. This mechanism is similar to the wave-breaking, when the velocity of some electrons participating in the wave becomes larger than the peak velocity of the wave itself. The electrons are injected each time the laser pulse gets focused, which is seen in Fig. \ref{secondFig}(a) and (b). There is a correlation between the evolution of the laser envelope and the spectrum of the injected electrons. The parameters of the presented simulation are: $n_0=2.5\cdot 10^{17}$ cm$^{-3}$, $r_c=1.43\lambda_p$, $a_0=2.6$, $w_0=0.72\lambda_p$, $b=4, \epsilon=1.5$ and the intensity envelope FWHM duration is $t_L=0.3\lambda_p/c=69$ fs. The envelope oscillation is damped due to broadening of the laser frequency spectrum, or red-shifting \cite{redshift}. Different wavelengths oscillate with different periodicity and they have different phase velocity, which leads to an overall smoothing of the oscillation \cite{EsareyGuide}.

\begin{figure}[h!]
\centering
\includegraphics[trim= 0 0mm 0mm 10mm,width=0.85\textwidth]{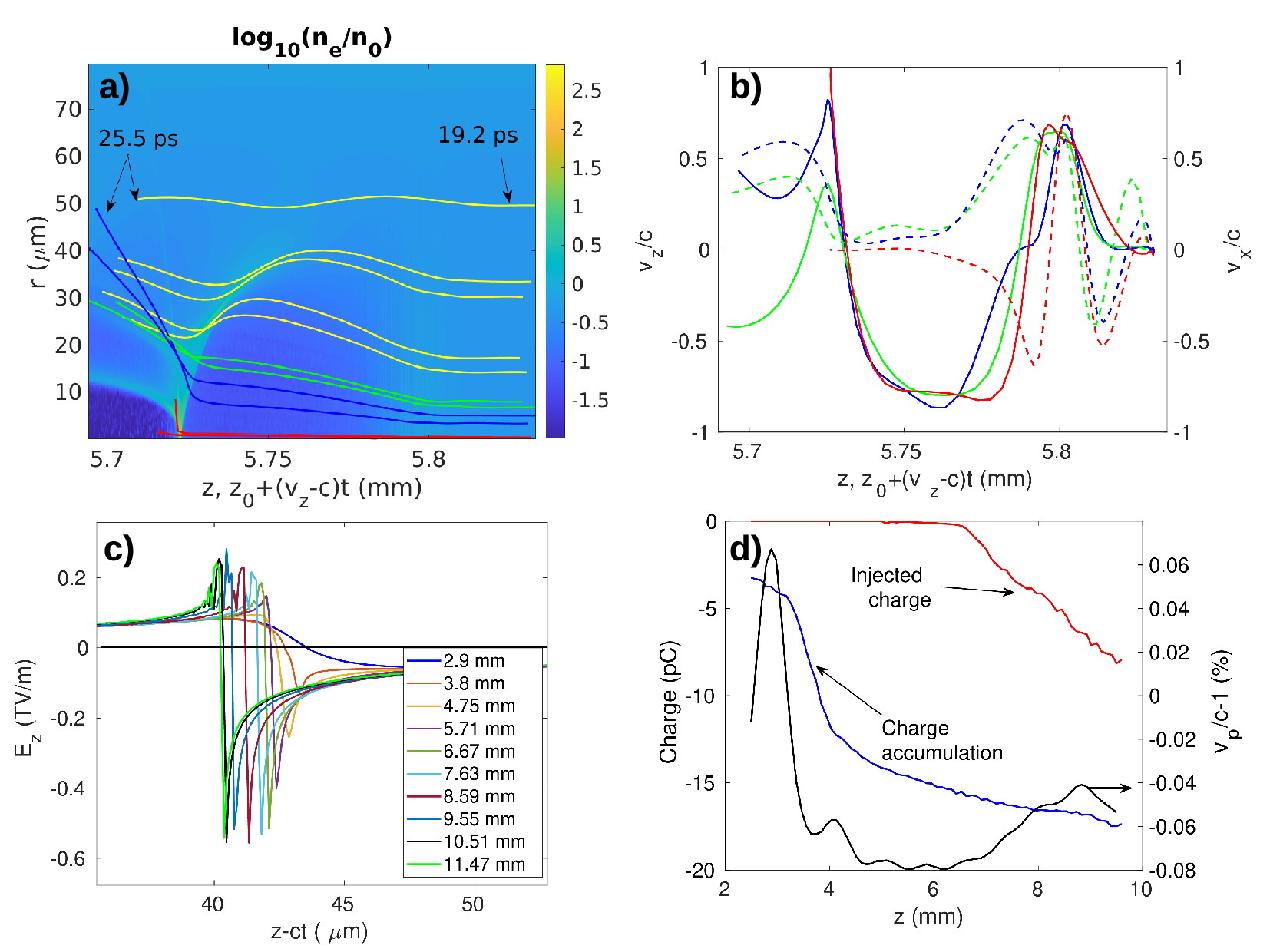}

\caption{{\bf Details of longitudinal self-injection:} (a) Electron density distribution overlaid with several electron trajectories originating from different radial positions. Here $z_0$ is the initial position of the tracked electrons and $v_z$ is their longitudinal velocity. Some of them, which are close to the propagation axis, can be injected (red curves). (b) Velocity phase-space evolution of selected electrons from different region of the plasma. The colors correspond to the electrons shown by the same colors in (a). The right vertical axis shows the transversal velocity (dashed lines). The red trajectory corresponds to an electron which gets injected. (c) Temporal evolution of the axial accelerating field in the moving frame. (d) Evolution of the charge in the back of the bubble (blue) and of the injected charge (red). The black curve (right axis) shows the phase velocity of the density peak.  }
\label{thirdFig}
\end{figure}

The variation in the laser peak intensity is imprinted in the evolution of the cavity length, that is shown in Fig. \ref{secondFig}(c). The effect of self-phase modulation \cite{selfCompr1, selfCompr2} is clearly visible in the density modulation near the position of the laser pulse (indicated by the white arrow), which gets sharper due to the self-compression. However, this effect appears only at the end of the acceleration. At the beginning of the propagation the oscillation of the bubble leads to a few microns variation in the position of the bubble's peak density in a time window of $(10$ mm$)/c\approx 30$ ps, which results in $\sim 10^{-4}$c oscillation amplitude of the bubble's phase velocity ($v_p$). This variation of the phase velocity leads to the self-injection of the electrons accumulated in the back of the bubble, where they experience $\sim 0.1$ TV/m accelerating field. In a distance of 1 mm some part of the electrons can easily gain $>100$ MeV energy, corresponding to a Lorentz factor $\gamma_e > 200$, that is much larger than the Lorentz factor associated with the phase velocity $\gamma_p=(1-v_p^2/c^2)^{-1/2}\approx 100$. As we show in the following, the oscillating phase velocity leads to an oscillation in the space charge field near the back of the bubble, which is the main reason for self-injection to occur.

One can understand the process of self-injection by taking a closer look at the kinematics of a single electron moving in an evolving bubble. From Fig \ref{thirdFig}(a) it is clear that electrons originating from a region close to the axis ($r=0$) can be injected (red trajectories). Electrons being farther away from the axis experience a transverse ponderomotive force, due to the Gaussian laser envelope, and contribute to the formation of the wake structure. By looking at the velocity components of these electrons (in Fig. \ref{thirdFig}(b)) one can see that their maximum velocity does not approach the speed of light, but it is large enough to spend some time in the back of the bubble where the high density shell is formed. The highest density is reached on the axis (density peak), where the electrons have the highest velocity (close to the speed of light) and they can contribute for a longer time to the charge accumulated in this region, which is practically co-moving with the laser. The charge accumulation is clearly visible in Fig. \ref{thirdFig}(d), and it is evident that electrons can spend more time in the density peak when the phase velocity of the wake is less than the speed of light (see the black curve in Fig. \ref{thirdFig}(d)). The charge in the back of the bubble is calculated by summing up all electrons that are in the vicinity of the density peak, within $4\,\mu$m distance. When the accumulated charge and the accelerating field are high enough (at the position $\approx 7$ mm in Fig. \ref{thirdFig}(c)) the electrons can acquire a velocity that is large enough for wave-breaking (red curve in Fig. \ref{thirdFig}(d)). This mechanism is called longitudinal self-injection \cite{longInj}.

\begin{figure}[h]
\centering
\includegraphics[trim= 0 0mm 0 0mm,width=0.4\textwidth]{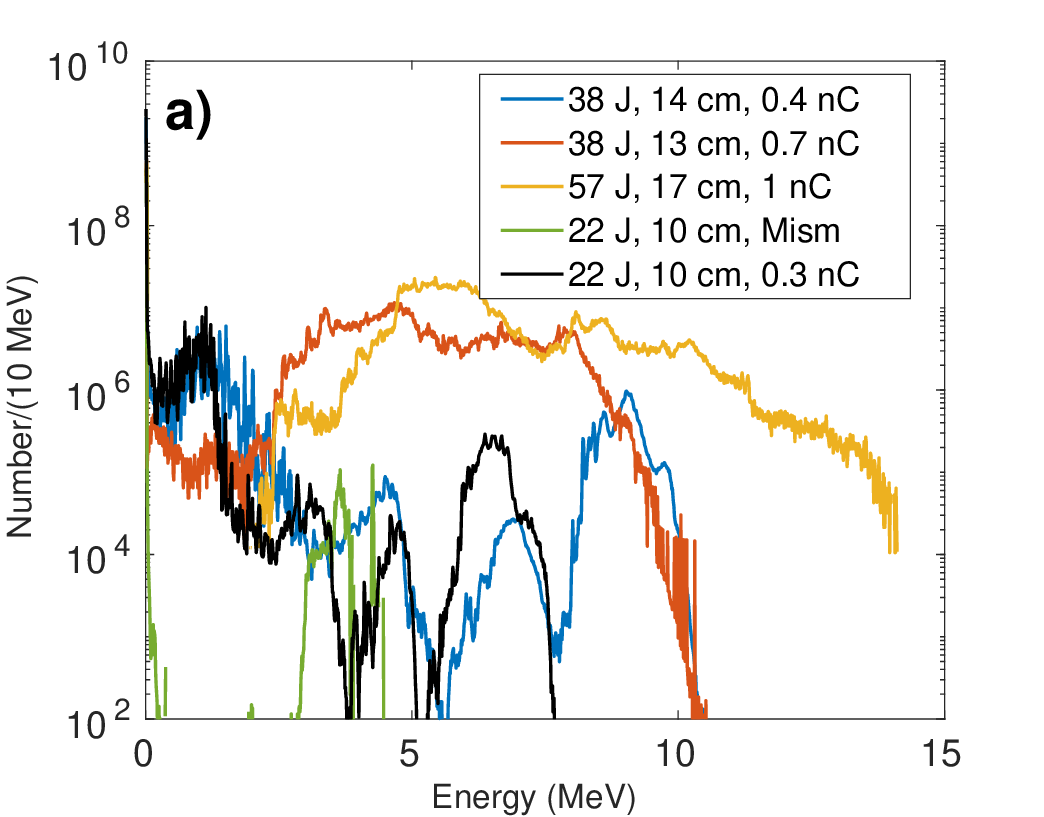}
\includegraphics[trim= 0 0mm 0 0mm,width=0.4\textwidth]{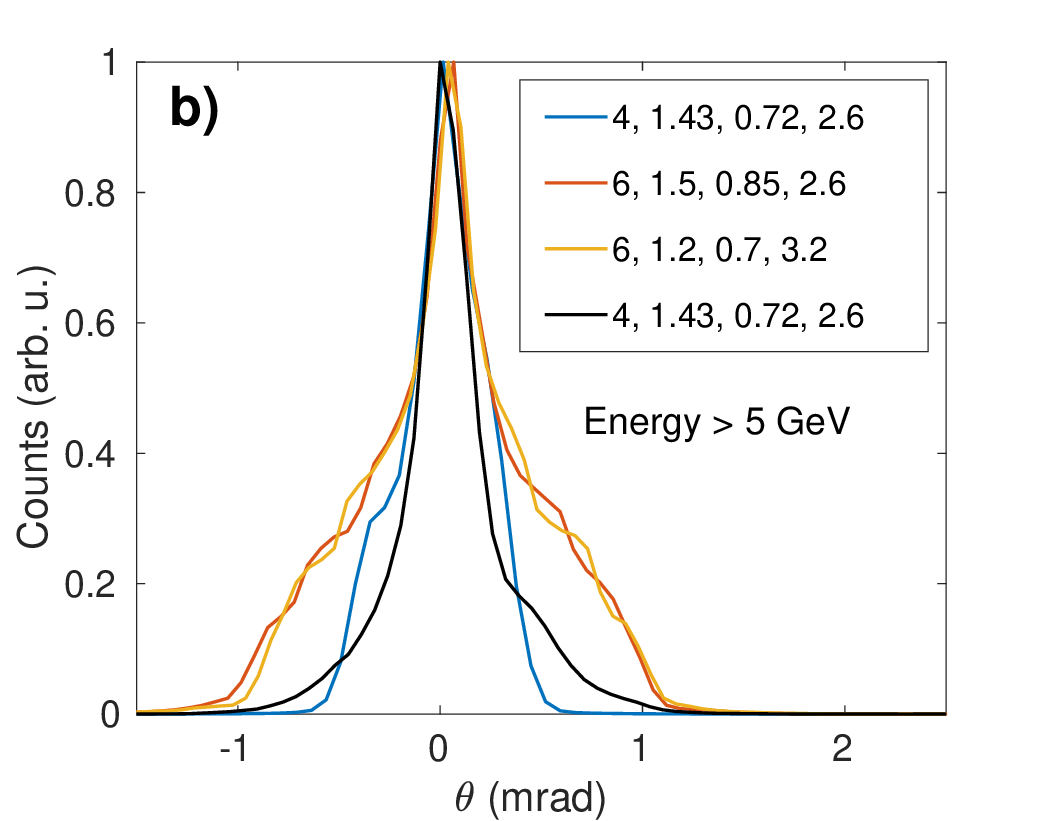}

\caption{{\bf Electron beam properties for different laser and plasma parameters:} (a) Energy spectra for different laser energies and channel lengths. (b) The corresponding divergence angle of electrons in the plane of laser polarization ($\theta=\arctan(p_x/p_z)$). The numbers in the legend represent the following parameters: $b, r_c/\lambda_p, w_0/\lambda_p, a_0$. }
\label{spectraComp}
\end{figure}

The concept of wide plasma channel was tested for different laser energies and geometrical parameters, that are summarized in Fig. \ref{spectraComp}(a). These results suggest that the maximum electron energy can be well above 10 GeV using a PW-class laser pulse and a 14 cm long channel. An interesting fact is that even after this significant energy gain the laser pulse still contains 70 \% of its initial energy, therefore in the case of the blue curve in Fig. \ref{spectraComp}(a) a short high-density gas jet can be placed after the 14 cm long channel to boost the electron energy even more. It is also seen that a mismatched laser pulse generates smaller cut-off energy and much lower injected charge compared to the mode-matched case. The mismatched case (green curve in Fig. \ref{spectraComp}(a)) corresponds to the LWFA in a narrow channel, which was realized experimentally \cite{OFI}, where no self-injection was observed, electrons were accelerated when ionization injection was used \cite{modeEvol}. The fact that in our simulation we observe self injection is attributed to the steepness of the channel density, which was $b=4$ in this case. With $b=6$ we did not observe self-injection in this mismatched case. The irregular envelope oscillation (see Fig. \ref{Compare}) caused by the presence of higher order modes prohibits the charge accumulation, shown in Fig. \ref{thirdFig}(d), thus the probability of electron self-injection is much smaller.

The immediate consequence of longitudinal self-injection is the low transverse emittance of the accelerated electron bunch. From Fig. \ref{thirdFig}(a),(b) it is clear that the injected electrons have a very small transverse momentum and a micrometer-scale transverse displacement which leads to small divergence and source size, which is shown in Fig. \ref{spectraComp}(b). The mismatched case is not shown here, because we considered electrons with higher than 5 GeV. 

The robustness of the wide channel LWFA was further tested with 15 J pulse energy as well, which was used in experiments \cite{OFI}. For $a_0\approx 2.5$ the required laser spot size is around $w_0=34\,\mu$m and the matched plasma density and channel radius are $5\times 10^{17} $cm$^{-3}$ and $r_c=60\,\mu$m, respectively. The simulation showed that, similarly to the experiment, 5 GeV electron cut-off energy is reached, but the acceleration distance was only 7 cm, i. e. 3 times shorter than that used in the experiment.

\section{Conclusions}

We presented a yet unexplored regime of wide channels, where  $r_c>lambda_p>w_0$ and $a_0>2$. This regime of laser guiding has been analyze and explain with three different methods: analytical, numerical (SDE), and fully self-consistent relativistic PIC modeling. In this regime the development of higher order modes can be greatly suppressed that enhances the LWFA efficiency and eliminates the problem of leaky channels and mode dispersion. Furthermore, we prove that the mode matching is continuously satisfied in the channel even if the laser spot oscillates because the mode matching does not require exact conditions, the parameters can vary within ~ 10 \% interval.

Due to the laser envelope  oscillations electrons can get self-injected from the background plasma, but at lower gas densities ionization injection can be also used, which allows a more precise control over the position of electron injection. The energy gain presented here fits well, and even surpasses, the theoretical prediction of the dephasing-limited LWFA (Eq. 6 in Ref. \cite{WLu}), while in recent experiments the maximum energy barely reaches \cite{gonsales}, or it is lower than the theoretical estimate \cite{OFI, modeEvol}. If we consider the mentioned energy scaling and 2 PW laser power, it will be possible to reach 40 GeV electron energy with an axial plasma density of $3.3\times 10^{16}$ cm$^{-3}$ and with a laser spot radius of $120\,\mu$m. In this case the required laser pulse energy is 200 J (pulse duration is 100 fs) and the acceleration distance is slightly longer than 1 meter \cite{meterScale}, that scales with the Rayleigh length (or with square of the laser spot size). This energy is already very close to the maximum electron energy ever produced in linear accelerators \cite{slac}.

\section*{Appendix}

\subsection*{Main parameters of a HOFI plasma channel}

The radial density profile in discharge and hydrodynamic plasma channels are commonly described by a polynomial function, which is multiplied by a Super-Gaussian in order to mimic the density modulation inside the radially expanding hydrodynamic shock:

\begin{equation}\label{eq:densprof}
n_e=n_0(1+\epsilon \rho^b)\exp(-\beta \rho^p),
\end{equation}
where $\rho=r/r_c$, $\beta\ll 1$, $p\gg 1$ and $b$ is a positive number characterizing the channel density profile. The characteristic radius of the channel is $r_c$. The parameter $\beta$ is easily expressed as function of the other parameters by defining the maximum density in the channel wall $n_e(\rho_m)=n_{max}$, where typically $n_{max}\sim 20 n_0$ and $\rho_m\approx 2$ is the position of the maximum density. The value of $\rho_m$ can be varied and it directly depends on the channels' expansion time (or on the delay of the main pulse). The values $\epsilon$ and $b$ can be related by the requirement $dn_e(\rho_m)/dr=0$, leading to the equation:

\begin{equation}\label{eq:epsilon}
\beta p\rho_m^{p-1}(1+\epsilon \rho_m^b)=\epsilon b \rho_m^{b-1},
\end{equation}
where $\beta=-(1/\rho_m^p)\ln[(n_{max}/n_0)/(1+\epsilon \rho_m^b)]$. This equation is solved numerically to obtain the set of parameters that are used in our analysis presented in the main text. In the Table below $n_{max}=16n_0$.

\begin{table}[h]
\centering
\includegraphics[trim= 0 30mm 0 0mm,width=0.8\textwidth]{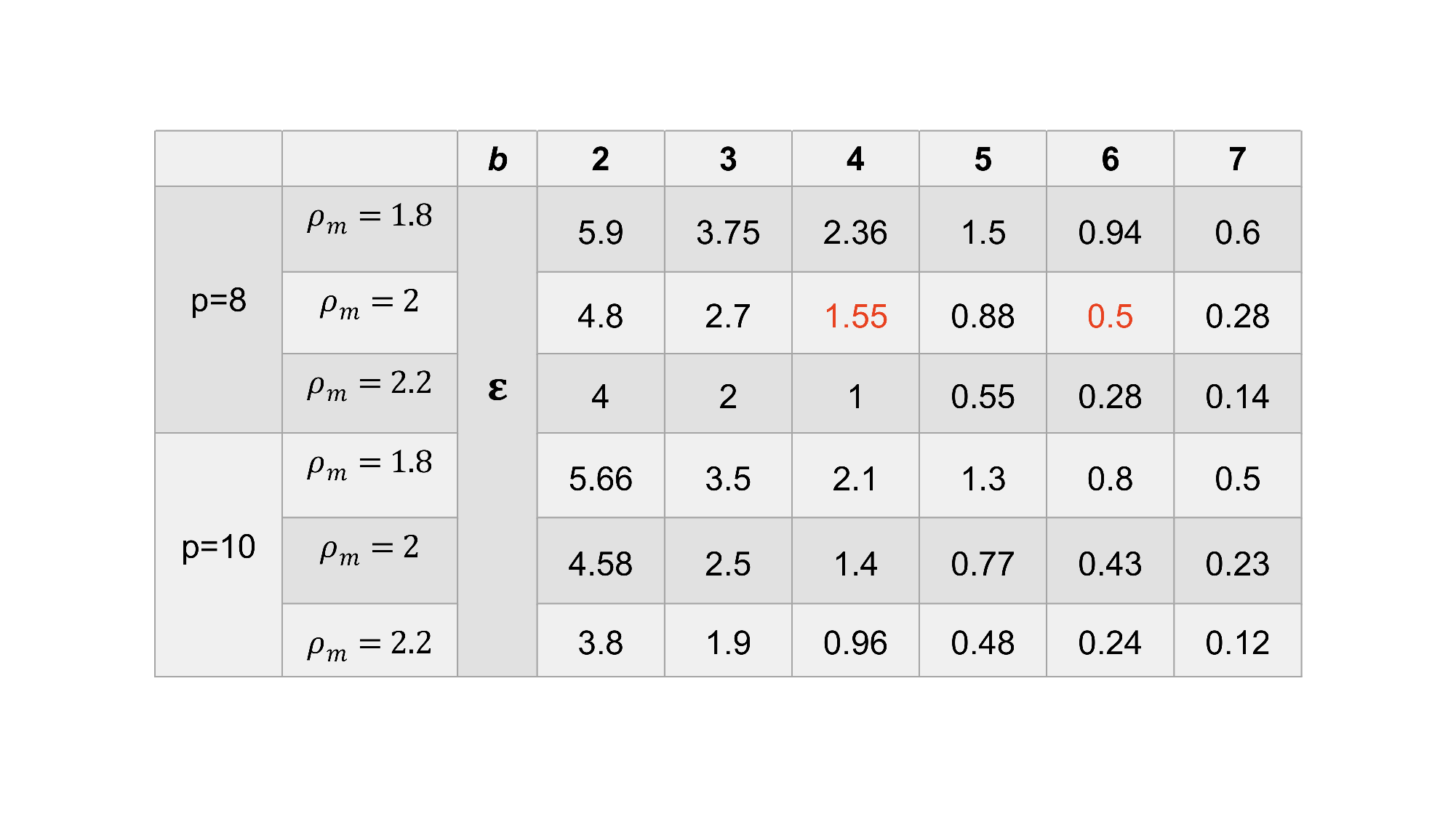}

\caption{Parameters of the plasma channel: Values of $\epsilon$ (channel wall steepness) for different $p$, $b$ and $\rho_m$ parameters. The red numbers are the values used in or simulations.}
\label{Table1}
\end{table}

\subsection*{Source-dependent expansion method for analysis of radial mode evolution}

Here we recall the main equations of the source-dependent expansion method, developed in Ref. \cite{SDE}, with the assumption of full rotational symmetry, meaning that the higher-order azimuthal modes are neglected. For the modeling of the spatial (or temporal) evolution of the laser envelope we consider the reduced wave equation, Eq. \ref{eq:helm}:

\begin{equation}\label{eq:paraxial}
\nabla_{\perp}^2 {E_L} +2 i k \partial {E_L}/\partial z = S = k^2[n_e/(\gamma n_{c})]{E_L}=k^2\chi {E_L},
\end{equation}
where the higher order longitudinal derivatives are neglected which corresponds to the slowly varying envelope approximation. The next step is to express the solution of Eq. (\ref{eq:paraxial}) as a combination of Laguerre-Gaussian modes:

\begin{equation}\label{eq:LGcomb}
\frac{E_L}{E_0}=\sum_{m=0}^M c_m D_m=\sum_{m=0}^M c_m L_m^0(2r^2/w^2) \exp[-(1-i \alpha)r^2/w^2],
\end{equation}
where $\alpha$ is related to the wavefront curvature and $c_m$ are the mode coefficients, which are complex numbers containing information about the phase shifts and they are proportional to $w_0/w$. In the case of the fundamental mode, in vacuum, $\alpha=z/z_R$, where $z_R=\pi w_0^2/\lambda$ is the Rayleigh length and $w=w_0\sqrt{1+(z/z_R)^2}$, which are used as initial conditions in the presented model, assuming that the focus position is at $z=0$. By substituting Eq. (\ref{eq:LGcomb}) into Eq. (\ref{eq:paraxial}) it is possible to derive a simple equation for the evolution of the mode coefficients \cite{SDE}:

\begin{equation}\label{eq:amdiff}
\partial c_m/\partial z + A_m c_m - i m B c_{m-1} - i(m+1) B^* c_{m+1} = -i F_m,
\end{equation}
where $B=F_{1}/c_{0}$, and the superscript $*$ refers to the complex conjugate and $A_m$ has the form:

\begin{equation}\label{eq:Am}
A_m(z)= 2\alpha/(k w^2) - (B)_i + i(2m+1)[2/(k w^2) + (B)_r],
\end{equation}
where the subscripts refer to the real ($()_r$) and imaginary ($()_i$) part of the quantity. 

The function $F_m(z)$ represents a coupling term between the neighboring radial modes and it is calculated using the source term:

\begin{equation}\label{eq:Fm}
F_m(z)=\frac{2}{k} \int_0^{\infty} (r/w^2) S D_m^* dr.
\end{equation}

Finally the evolving parameters $\alpha, w$ are calculated with the following equations:

\begin{equation}\label{eq:wdiff}
w' = 2\alpha/(k w) - (B)_i w,
\end{equation}

\begin{equation}\label{eq:alphadiff}
\alpha' = 2(1+\alpha^2)/(k w^2) + 2[(B)_r - (B)_i \alpha],
\end{equation}

Equations (\ref{eq:amdiff} - \ref{eq:alphadiff}) can be used to calculate the beam profile at any position $z$ using Eq. (\ref{eq:LGcomb}) with the assumption that $m=0$ is the strongest mode and $c_{0}$ is the largest amplitude in the field composition. This condition holds in the case of close-to-matched beam profiles, which we consider in our work.

\hspace{20mm}

{\bf Acknowledgement}

We acknowledge KIF\"U/NIIF for awarding us access to the Komondor HPC located in Debrecen, Hungary. The ELI-ALPS project (GINOP-2.3.6-15-2015-00001) is supported by the European Union and co-financed by the European Regional Development Fund. 

\hspace{10mm}

{\bf Data availability statement}
The data that support the findings of this study are available from the corresponding author upon reasonable request.



\end{document}